\documentstyle[12pt]{article}
\def\be{\begin{equation}}
\def\ee{\end{equation}}
\def\ba{\begin{eqnarray}}
\def\ea{\end{eqnarray}}
\def\d{D^0}
\def\db{\bar{D}^0}
\def\mixing{$\d$--$\db$ mixing}
\def\system{$\d$--$\db$ system}
\def\espaco{$\d$--$\db$ space}
\def\ddbar{$\d$/$\db$}
\def\a{\bar{A}}
\def\b{\bar{B}}
\def\cx{\cal{X}}
\def\A{|A|^2}
\def\B{|B|^2}
\def\Ab{|\bar A|^2}
\def\Bb{|\bar B|^2}
\title{\bf CP-Violating Observables in Tagged \\
\ddbar\ Decays}
\author{G.\ C.\ Branco,
W.\ Grimus\thanks{On leave of absence
(from 1 October 1995 to 31 January 1996)
from the University of Vienna,
Institute for Theoretical Physics,
Boltzmanngasse 5, A-1090 Vienna}\ \
and L.\ Lavoura\thanks{Researcher of the
Technical University of Lisbon}  \\
\\
\small CFIF, Instituto Superior T\'ecnico,
Edif\'\i cio Ci\^encia (f\'\i sica) \\
\small P-1096 Lisboa Codex, Portugal}
\begin{document}
\maketitle
\begin{abstract}
We perform a systematic study of CP-violating observables
in \ddbar\ decays.
We show that given a final state,
which may be a CP eigenstate or not,
all the observables can in principle be measured
if decay-time information is available and if,
for CP non-eigenstates,
both decay rates into $f$ and $\bar f$ are considered.
As an illustration
of our analysis
we discuss an effective superweak scenario.
\end{abstract}


\section{Introduction}
The study of mixing and decays of neutral $D$ mesons
is an important testing ground
for the Standard Model (SM),
with the potential for discovery of new physics.
(See,
for instance,
refs.~\cite{big, bur, liu}.)
In the SM,
short-distance contributions to mixing
arise from the box diagrams \cite{dat},
and are negligible \cite{gol}:
$\Delta m_D^{\mbox{\scriptsize box}} (SM) \approx 10^{-17}$ GeV.
Initially,
it was thought that long-distance effects
arising from second-order weak interactions with mesonic intermediate states
would give a much larger contribution,
about two orders of magnitude larger than the short-distance one
\cite{wol1}.
It was later argued by Georgi \cite{geo}
that cancellations occur among the various dispersive channels,
making the overall long-distance contribution
smaller than initially estimated.
Recently,
this conjecture has been confirmed
by an analysis of \mixing\ in the framework of
heavy quark effective field theory (HQEFT)
including leading order QCD corrections \cite{ohl},
which gave $\Delta m_D^{\mbox{\scriptsize HQEFT}} (SM)
\approx (0.9-3.5)\cdot 10^{-17}$ GeV.
One concludes therefore that the observation of \mixing\ at
any existing or planned facility
would be evidence for physics beyond the SM.

The study of \mixing\ also provides the exciting possibility
of  detecting new sources of CP violation beyond the
Kobayashi-Maskawa (KM) mechanism of the SM.
Indeed,
most of the models beyond the SM
which predict significant contributions to \mixing\
contain new sources of CP violation. Recently, it has been emphasized that
CP violation can play an important r\^{o}le
in the study of \mixing\
\cite{wol2, bla, bro}.

In this paper we perform a systematic study of
a class of
physical observables
which can provide evidence for CP violation in the \system.
We then assume
(as was also done in refs. \cite{wol2, bro})
that the dominant new source of CP violation
in the \system\ arises from
an effective superweak interaction (ESI) leading
(in a suitable phase convention)
to a non-vanishing $\mbox{Im} \: M_{12}$,
while $\Gamma_{12}$ and the decay amplitudes are dominated
by the SM contributions,
and are real (we neglect small phases in the KM matrix
proportional to $\lambda^4$ in the Wolfenstein parameterization).
Although this is a specific assumption,
it turns out that a large number of models
which can lead to significant \mixing\ belong to the ESI class
\cite{bra}.
Our analysis is complementary to the work of ref.~\cite{ley},
where an opposite assumption was made,
namely that physics beyond the SM contributes significantly
only to the decay amplitudes.

We will start by identifying CP-violating observables in tagged
\ddbar\ decays to specific final states,
like CP non-eigenstates $f/\bar f$,
and CP eigenstates $F$.
We do this by studying the behaviour
of various quantities under phase redefinitions of the state vectors.
We point out that
the study of the time-dependent decay rates of $\d$ and $\db$
to a final state $f$ {\em and}
its antiparticle state $\bar f$ can,
at least in principle,
allow the measurement of all the CP-violating observables
which we have identified.
The same is true in the case of decays to CP eigenstates.
Furthermore,
after having switched to an ESI scenario,
we consider $f=K^-\pi^+ /\bar f=K^+\pi^-$ and
$F=K^+K^-$ as illustrative examples
and possible candidates for an experimental search for CP violation.

\section{Phenomenology of CP-violation in tagged \ddbar\ decays}
We want to describe the decays of $\d$ and $\db$
to a final state $|f>$ and its antiparticle state $|\bar f>$.
We first assume that $|f>$ is not a CP eigenstate.
Phenomenologically,
there are four independent amplitudes
\be
\cal{A}(\d \rightarrow f) \equiv A\, , \
\cal{A}(\db \rightarrow \bar{f}) \equiv \a\, , \
\cal{A}(\db \rightarrow f) \equiv B\, , \
\cal{A}(\d \rightarrow \bar{f}) \equiv \b\, ,
\label{eq:amplitudes}
\ee
entering in the description of those decays.
The mass eigenstates of the charmed neutral mesons are
\ba
| D_H > & = & p |\d> + q |\db>\, ,
\nonumber\\
| D_L > & = & p |\d> - q |\db>\, ,
\label{eq:dhdl}
\ea
with $ |p|^2 + |q|^2 = 1 $ and
\be
\frac{q}{p} =
\frac{\Delta m - \frac{i}{2} \Delta \Gamma}{2 (M_{12}
- \frac{i}{2} \Gamma_{12})} =
\frac{2 (M_{12}^* - \frac{i}{2} \Gamma_{12}^*)}{\Delta m
- \frac{i}{2} \Delta \Gamma} =
\sqrt{\frac{
M_{12}^* - \frac{i}{2} \Gamma_{12}^*
}{
M_{12} - \frac{i}{2} \Gamma_{12}}}\, ,
\label{eq:qp}
\ee
where $\Delta m = m_H - m_L > 0$
($H$=heavy,
$L$=light),
$\Delta \Gamma = \Gamma_H - \Gamma_L$,
and $M_{12} - \frac{i}{2} \Gamma_{12}$
is the off-diagonal matrix element in the effective time evolution
in the \espaco.
The sign of the square root in eq.~(\ref{eq:qp})
is chosen such that $\Delta m > 0$.

In quantum mechanics,
physical quantities cannot depend on the phases
of the initial and final states.
In our case,
physics is therefore invariant under the phase redefinitions
\be
|\d> \rightarrow e^{i \gamma} |\d>\, , \
|\db> \rightarrow e^{i \bar{\gamma}} |\db>\, , \
|f> \rightarrow e^{i \gamma_f} |f>\, , \
|\bar{f}> \rightarrow e^{i \bar{\gamma}_f} |\bar{f}>\, ,
\label{eq:redefinitions}
\ee
with four independent phases $\gamma$,
$\bar{\gamma}$,
$\gamma_f$ and $\bar{\gamma}_f$.
Noting that under such a redefinition
\be
A \rightarrow e^{i(\gamma - \gamma_f)} A \, , \
B \rightarrow e^{i(\bar \gamma - \gamma_f)} B \, , \
\a \rightarrow e^{i(\bar \gamma - \bar \gamma_f)} \a \, , \
\b \rightarrow e^{i(\gamma - \bar \gamma_f)} \b \, , \
\frac{q}{p} \rightarrow e^{i(\gamma - \bar \gamma)} \frac{q}{p}\, ,
\label{reph}
\ee
we see that the
only quantities which are at most quadratic in the amplitudes
and which have a chance to be physical are
\be
\label{phys}
\eta , \
\A , \
\Ab , \
\B , \
\Bb , \
\frac{q}{p} A^* B , \
\frac{p}{q} \a^* \b , \
\frac{q}{p} \a \b^* , \
\frac{p}{q} A B^* ,
\ee
where $\eta \equiv |q/p|$.
Because of the possibility of the phase redefinitions (\ref{reph}),
there are only seven independent
quantities in eq.~(\ref{phys}),
five moduli and two phases.
One could,
for instance,
eliminate $\B$,
$\Bb$,
$(q/p) \a \b^*$ and $(p/q) A B^*$
from the set in eq.~(\ref{phys}) and take
$\eta , \
\A , \
\Ab , \
|(q/p) A^* B| , \
|(p/q) \a^* \b| , \
\arg ((q/p) A^* B)$ and
$\arg ((p/q) \a^* \b) $
as a set of independent quantities. Note that instead of $\eta$ very often
the phase $\arg (\Gamma_{12}^* M_{12})$ is used, since $\eta$ is a function
of this phase and of $\Delta m, \, \Delta \Gamma$. However, in the following
discussion this question of independence will not be important.

In order to study CP violation
we first consider the consequences of CP invariance.
If CP is a good symmetry,
there exist two phases $\beta$ and $\beta_f$,
and a CP eigenvalue $\zeta_D$ ($\zeta_D^2 = 1$),
such that
\ba\label{CPtrafo}
\cal{CP} |\d > = e^{i\beta} |\db >\, ,
& &
\cal{CP} |\db > = e^{-i\beta} |\d >\, ,
\nonumber\\
\cal{CP} |f> = e^{i\beta_f} |\bar f>\, ,
& &
\cal{CP} |\bar f> = e^{-i\beta_f} |f>\, ,
\ea
and
\ba
\cal{CP} |D_H> & = & \zeta_D |D_H>\, ,
\nonumber\\
\cal{CP} |D_L> & = & - \zeta_D |D_L>\, .
\ea
As a result,
the CP invariance conditions
are
\be \label{CPcond}
\a = e^{i(\beta_f - \beta)} A , \
\b = e^{i(\beta_f + \beta)} B , \
\frac{q}{p} = \zeta_D e^{i\beta} , \
M_{12} = e^{-2i\beta} M_{12}^* \, , \
\Gamma_{12} = e^{-2i\beta} \, \Gamma_{12}^* \, .
\ee
Therefore,
if CP is conserved the quantities
\be
\eta^2 - 1 , \
\A - \Ab , \
\B - \Bb , \
\cx \equiv \frac{q}{p}A^*B - \frac{p}{q}\a^*\b \, , \
\bar{\cx} \equiv \frac{q}{p}\a \b^* - \frac{p}{q}AB^*
\label{CPcons}
\ee
all vanish.
On the other hand,
if CP is not conserved the quantities in
eq.~(\ref{CPcons}) measure CP violation.
Knowing their values means having
the complete information about CP violation
in \mixing\ and in the decays of \ddbar\
into the specific final states $f/\bar f$.
Furthermore, all
observables in this connection must be functions of the quantities
in the set of eq.~(\ref{CPcons})
and possibly of
\be
\A + \Ab , \
\B + \Bb , \
\cal{Y} \equiv \frac{q}{p} A^*B + \frac{p}{q} \a^*\b \, , \
\bar{\cal{Y}} \equiv \frac{q}{p}\a \b^* + \frac{p}{q}AB^* ,
\ee
which are not CP-violating. CP-violating observables have to be odd functions
of the quantities in eq.~(\ref{CPcons}).

It is easy to see that $\bar{\cx}$
is related to $\cx$ and $\cal{Y}$ by
\be
\bar{\cx} =
- \frac{1}{2} (\eta^2+\frac{1}{\eta^2}) \cx^*
+ \frac{1}{2} (\eta^2-\frac{1}{\eta^2}) \cal{Y}^* .
\label{x=x}
\ee
As we will see below,
instead of using
$\mbox{Re} \: \cx$ and $\mbox{Im} \: \cx$
(or $\mbox{Re} \: \bar{\cx}$ and $\mbox{Im} \: \bar{\cx}$)
as independent CP-violating observables,
it is more convenient to take
\ba
\mbox{Im} \: [(\Delta m-\frac{i}{2}\Delta \Gamma) \cx ]
& = &
- 2\, \mbox{Im} \:
[M_{12} (AB^*+\a^*\b )] - \mbox{Re} \: [\Gamma_{12}(AB^*-\a^*\b )] \, ,
\nonumber\\
\mbox{Im} \: [(\Delta m-\frac{i}{2}\Delta \Gamma) \bar{\cx} ]
& = &
- 2\, \mbox{Im} \:
[M_{12} (AB^*+\a^*\b )] + \mbox{Re} \: [\Gamma_{12}(AB^*-\a^*\b )] \, .
\label{im}
\ea
It can be checked by using eq.~(\ref{x=x})
that  both sets are equivalent
if $\Delta m \Delta \Gamma \neq 0$.
To derive eqs.~(\ref{im}) we have used eq.~(\ref{eq:qp}).

It is important to realize
that there are three possible different kinds of CP violation
in the \system\
(or any other neutral-meson system).
Namely,
there may be CP violation in mixing
(which just means $\eta \neq 1$),
direct CP violation
(that is,
$ |A| \neq |\bar A|$ or $ |B| \neq |\bar B|$),
or CP violation in the relationship between mixing and decays
--- non-vanishing $\cal{X}$ or $\bar{\cal{X}}$.
Often, authors refer to this last form of CP violation as
``CP violation in mixing'',
but we think this designation is inappropriate.
CP is violated in mixing if and only if $\eta$ is different from 1,
and this is independent of the decay amplitudes
to any particular decay modes.

If at $t=0$ a $\d$ or a $\db$ is produced,
then its time evolution is given by
\ba
|\d (t)> & = & f_+(t) |\d > + \frac{q}{p} f_-(t) |\db > \, ,
\nonumber\\
|\db (t)> & = & \frac{p}{q} f_-(t) |\d > + f_+(t) |\db > \, ,
\ea
with
\be
f_\pm (t) = \frac{1}{2} e^{-i(m - \frac{i}{2}\Gamma)t}
\left[
e^{-\frac{i}{2}(\Delta m - \frac{i}{2}\Delta \Gamma)t} \pm
e^{\frac{i}{2}(\Delta m - \frac{i}{2}\Delta \Gamma)t}
\right],
\ee
where
\be
m = \frac{1}{2} (m_H + m_L)\, , \;
\Gamma = \frac{1}{2} (\Gamma_H + \Gamma_L)\, .
\ee
These formulae allow us to consider the time-dependent decay rates of \ddbar.
Now we address ourselves to the question of whether one can measure,
at least in principle,
all the CP-violating observables in the set (\ref{CPcons}),
which we derived from rephasing invariance,
by considering the above-mentioned time-dependent decay rates.
The answer is affirmative.
In order to see this,
consider the quantities \cite{bla}
\ba
\label{diffrate}
\lefteqn{\Gamma(\d (t) \rightarrow f)-\Gamma(\db (t) \rightarrow \bar f)
= e^{-\Gamma t}\,
\{
\A - \Ab + (\Gamma t)\, \mbox{Im} \: [(x-iy)\cx]+}
\nonumber\\[2mm]
& + \frac{1}{4}\, (\Gamma t)^2
\left[ (\eta^2\B - \frac{1}{\eta^2}\Bb )(x^2+y^2) -
(\A -\Ab )(x^2-y^2) \right] + \cal{O}(t^3)\} \, ,
\nonumber\\[3mm]
\lefteqn{\Gamma(\d (t) \rightarrow \bar f)-\Gamma(\db (t) \rightarrow f)
= e^{-\Gamma t}\,
\{
\Bb - \B + (\Gamma t)\, \mbox{Im} \: [(x-iy)\bar{\cx}]+}
\nonumber\\[2mm]
& +\frac{1}{4}\, (\Gamma t)^2
\left[ (\eta^2\Ab - \frac{1}{\eta^2}\A )(x^2+y^2) -
(\Bb -\B )(x^2-y^2) \right] + \cal{O}(t^3)\} \, ,
\ea
with
\be
x\equiv \frac{\Delta m}{\Gamma}\, , \;
y \equiv \frac{\Delta \Gamma}{2 \Gamma}\, .
\ee
The terms with time dependence $e^{-\Gamma t}$ are $\A -\Ab$ and $\Bb -\B$,
whereas those with time dependence $(\Gamma t\, e^{-\Gamma t})$
coincide with the quantities in eq.~(\ref{im}).

Taking also into account the sum of the time-dependent decay rates,
the moduli of all the amplitudes are found
from the terms with time dependence $e^{- \Gamma t}$.
This would in principle allow the determination of $\eta^2$
from the $((\Gamma t)^2 \, e^{- \Gamma t})$ contributions
in eqs.~(\ref{diffrate}).
However,
this contribution can rather be used to get information on $x$ and $y$,
while $\eta$ is better extracted from
inclusive semileptonic decays of tagged \ddbar\ by
\be
\label{CPmix}
\frac{\cal{N}_- -\cal{N}_+}{\cal{N}_- +\cal{N}_+}
=
\frac{\eta^4 - 1}{\eta^4 + 1}
=
\frac{\mbox{Im} \: (\Gamma_{12}^*M_{12})}{|M_{12}|^2
+ \frac{1}{4}|\Gamma_{12}|^2} \, ,
\ee
where $\cal{N}_+\, (\cal{N}_-)$ is the number of positively (negatively)
charged leptons originating from $\db (t) \, (\d (t))$.
Of course,
with like-sign dilepton events
from correlated $\d \db$ decays $(\eta^4-1)/(\eta^4+1)$
can also be measured.

If we consider the decays to CP eigenstates $|F>$
the analysis becomes simpler.
We must then make the identification
\be
B \rightarrow \a, \: \b \rightarrow A
\ee
in the previous formulas. Then we have $\cx = \bar{\cx}$ and
\be
\label{F}
\mbox{Im} \: [(\Delta m-\frac{i}{2}\Delta \Gamma )\cx ] =
-4\, \mbox{Im} \: (M_{12}A\a^*) \, .
\ee
We note that eq.~(\ref{F}) does not depend on $\Gamma_{12}$
and the two eqs.~(\ref{diffrate}) become identical
in the case $\d / \db \rightarrow F$.
Instead of the quantity on the right-hand side of eq.~(\ref{F})
we could take $\mbox{Im} \: (\frac{q}{p}A^*\a)$ as the CP-violating
quantity which is zero if CP is conserved.
Indeed,
if CP is conserved,
there exists a CP eigenvalue $\zeta_F = \pm 1$ such that
$\cal{CP} |F> = \zeta_F |F>$,
with
\be
\a = \zeta_F \, e^{-i\beta}A \, .
\ee
Therefore,
\be
\frac{q}{p}A^*\a =\zeta_D \zeta_F \A\, ,
\quad M_{12}A\a^*=(M_{12}A\a^* )^*
\ee
are both real.

\section{The effective superweak interaction}
In the SM,
CP violation in the decays of \ddbar\ is extremely small.
This can easily be seen when using
the Wolfenstein parameterization of the KM matrix.
Going beyond the SM,
we assume that,
either we can neglect the new contributions
to the decay amplitudes and to $\Gamma_{12}$,
or the new contributions to the decay of the $c$ quark
are real relative to the SM ones.
On the other hand,
the additional interactions enhance $M_{12}=|M_{12}|e^{i\phi}$.
Then,
if the phase $\phi$ is large,
we have an effective superweak scenario
with $|M_{12}| \gg |\Gamma_{12}|$,
while \cite{wol2, bro} all the CP-violating quantities in eq.~(\ref{CPcons})
arise only from $\phi$.
Assuming for simplicity that $\Gamma_{12}$ is exactly real
and that there is no CP violation in $A$,
$B$,
$\a$,
and $\b$,\footnote{This amounts
to CP conservation in the interactions relevant for the decays,
with a phase convention $\beta=\beta_f=0$ in eq.~(\ref{CPtrafo}).}
we obtain
\be
\Delta m \approx 2|M_{12}| \, ,
\qquad
\Delta \Gamma \approx 2\Gamma_{12} \cos \phi \, ,
\ee
and
$$\A - \Ab = \B - \Bb = 0 \, ,$$
$$\mbox{Im} \: [(\Delta m - \frac{i}{2}\Delta \Gamma)\cx ] \approx
\mbox{Im} \: [(\Delta m - \frac{i}{2}\Delta \Gamma)\bar{\cx}] \approx
-2\, \mbox{Im} \: [M_{12}(AB^* + A^*B)] \, ,$$
\be
\label{ESI}
\frac{\eta^4-1}{\eta^4+1}
\approx
2 \frac{\Gamma_{12}}{\Delta m} \sin \phi \, .
\ee
Since in an ESI the phase $\phi$ is the only source of CP violation,
there must be relations among the quantities of eq.~(\ref{CPcons}),
as borne out by eq.~(\ref{ESI}).

Considering first CP violation in mixing,
we must take into account that eq.~(\ref{CPmix}) is somewhat misleading,
because it hides the fact that $\cal{N}_+$ and $\cal{N}_-$
are suppressed by mixing,
and really contain a factor $x^2+y^2 \approx x^2$.
Taking this into account,
we see that in order to detect CP violation in mixing
one has to overcome a  suppression factor $\Delta m \Delta \Gamma / \Gamma^2$.
In the SM we expect $\Delta \Gamma /\Gamma \sim \Delta m(SM)/\Gamma
\sim 10^{-5\pm 1}$;
therefore,
even with $x^2 \sim 10^{-2}$ at the present upper limit
and $\sin \phi \sim 1$,
it seems to be hard to see an effect.

Taking this crude argument into account,
we observe that with an ESI only
the terms linear in $t$ inside the braces of eq.~(\ref{diffrate})
are relevant:
\ba
\Gamma(\d (t) \rightarrow f)-\Gamma(\db (t) \rightarrow \bar f) & \approx &
\Gamma(\d (t) \rightarrow \bar f)-\Gamma(\db (t) \rightarrow f)
\nonumber\\
  & \approx &
-2 \, (\Gamma t \, e^{-\Gamma t})\, |AB| x \cos \delta \sin \phi \, .
\ea
The phase $\delta = \arg (AB^*)$
is induced by different final-state interactions in the amplitudes $A$ and $B$,
and is estimated to be \cite{bro} between $5^0$ and $13^0$.
Therefore,
$\cos \delta \approx 1$.
In the case of CP eigenstates $F$,
there is no such phase and
\be
\Gamma(\d (t) \rightarrow F)-\Gamma(\db (t) \rightarrow F) \approx
-2 \, (\Gamma t \, e^{-\Gamma t})\, \A \zeta_F \, x \sin \phi \, .
\ee

The best final-state candidates to search for mixing and CP violation
are probably $f=K^- \pi^+/\bar f =K^+ \pi^-$,
with branching ratios \cite{part}
$ \mbox{BR} (D^0 \rightarrow K^- \pi^+) = (4.01 \pm 0.14) \cdot 10^{-2} $
and $ \mbox{BR} (D^0 \rightarrow K^+ \pi^-)
= (3.1 \pm 1.4) \cdot 10^{-4}$, respectively,\footnote{See,
however,
the remarks to the latter value in refs.~\cite{liu, part}.}
and $F=K^+K^-$,
with $\zeta_F=1$ and $ \mbox{BR} (D^0 \rightarrow K^+K^-) =
(4.54\pm 0.29)\cdot 10^{-3} $.
The sensitivity of these final states to CP violation is nearly the same,
because
\be
\frac{\mbox{BR}(K^+K^-)}{\sqrt{\mbox{BR}(K^-\pi^+)\mbox{BR}(K^+\pi^-)}}
= 1.29 \pm 0.30 \, .
\ee
For $f=K^- \pi^+/\bar f =K^+ \pi^-$ the decay rates
are usually normalized to the number of Cabibbo-allowed decays.
We thus consider the distribution in decay time
$$
\frac{[\Gamma(\d (t) \rightarrow f) - \Gamma(\db (t) \rightarrow \bar f)]dt}
{\int_{0}^{\infty}[\Gamma(\d (t') \rightarrow f) + \Gamma(\db (t')
\rightarrow \bar f)]dt'} \approx
\frac{[\Gamma(\d (t) \rightarrow \bar f) - \Gamma(\db (t) \rightarrow f)]dt}
{\int_{0}^{\infty}[\Gamma(\d (t') \rightarrow f) + \Gamma(\db (t')
\rightarrow \bar f)]dt'} \approx
$$
\be
\approx - \left[ \frac{\Gamma(\d \rightarrow K^+\pi^-)}{\Gamma(\d
\rightarrow K^-\pi^+)} \right]^{1/2} x \cos \delta \sin \phi \, \Gamma t\,
e^{-\Gamma t} d(\Gamma t) \, .
\ee
Experimentally,
the bound on $x$ is roughly 1/10.
Therefore,
with $10^3-10^4$ Cabibbo-allowed decays,
or 10--100 doubly-Cabibbo-suppressed decays,
and with decay-time information,
it might be possible to constrain $x\sin \phi$
beyond the trivial constraint $|x\sin \phi |\leq x$.
The effect of CP violation with an ESI
in the determination of mixing
from $\Gamma(\d (t) \rightarrow \bar f)+
\Gamma(\db (t) \rightarrow f)$ has been studied in refs.~\cite{bla, bro}.
It was found that,
at the present level of accuracy,
it is included in the experimental errors \cite{bro}.

For $F=K^+K^-$ we use the normalization
\be
\frac{[\Gamma(\d (t) \rightarrow F) - \Gamma(\db (t) \rightarrow F)]dt}
{\int_{0}^{\infty}[\Gamma(\d (t') \rightarrow F) + \Gamma(\db (t')
\rightarrow F)]dt'} \approx -x\sin \phi \, \Gamma t\,
e^{-\Gamma t} d(\Gamma t).
\ee
{}~
Here we estimate that $10^2-10^3$ $K^+K^-$ events
would probably give a useful constraint on $x\sin \phi$.

\section{Conclusions}
We have done a systematic study of CP violation in tagged $\d /\db$ decays,
in which we have identified
all the rephasing-invariant observables which measure CP violation
in the decays to a final state $f$ and its antiparticle state $\bar f$,
or to a CP eigenstate $F$.
These quantities,
which parameterize the complete information on CP violation in this context,
are constructed from the mixing parameter $q/p$
and from the decay amplitudes,
and therefore depend on the specific final state.
We have shown that by studying the time-dependent decay rates
of \ddbar\ to $f/\bar f$ or to $F$ one can,
in principle,
measure all the CP-violating observables which we have identified.
The \system\ is unique
(see \cite{gol} for a nice presentation of this fact),
in comparison to the $K^0$--$\bar{K}^0$ and $B^0$--$\bar{B}^0$ systems,
in that $x$ and $y$ are both much smaller than one.
As a consequence,
one can expand the time-dependent decay rates
in terms of $(\Gamma t)^n \, e^{-\Gamma t} \; (n=0,1,\dots )$,
and it is probably sufficient
to keep only terms with $n \leq 2$.
It is crucial for our analysis
that experimentally the coefficients of the terms with $n \leq 2$
can be extracted.

Although our analysis is model-independent,
we have considered the special case of the ESI hypothesis,
which assumes that physics beyond the SM does not
contribute significantly to the \ddbar\ decay amplitudes,
but gives the dominant contribution to $M_{12}$,
with an unsuppressed CP-violating phase
$\phi \equiv \arg M_{12}$.
Under this assumption,
we have tried to assess the sensitivity of experiments
with decay-time information to CP violation.
Thereby we have compared CP non-eigenstates
(where one has Cabibbo non-suppressed
versus doubly-Cabibbo-suppressed amplitudes)
with CP eigenstates
(single Cabibbo suppression).
We have seen that experimental information on CP violation restricts
(or determines)
the quantity \cite{wol2} $x\sin \phi$,
and the upper limit obtained for this quantity
must be smaller than the upper limit on $x$,
in order to get a non-trivial restriction in the $x-\phi$ plane.
In the case of $x$ close to its upper bound,
we have estimated the numbers of events necessary for this purpose.
They seem to be well within the range of current and future search
for \mixing\ \cite{liu}.
Of course,
an experimental determination of $x\sin \phi$
would not only prove the existence of CP violation
in the \system,
but also give a lower bound on the mixing parameter $x$.

\end{document}